\documentclass[14pt]{article}

\usepackage[top=2.0cm, bottom= 2.0cm, right=3.0cm, left=3.0cm, headheight=14pt]{geometry}
\usepackage[citestyle=numeric, sorting=none, natbib=true, backend=biber]{biblatex}

\bibliography{/home/sawy/postdoc_auc/strangeness_enhancement_paper/references}
\usepackage{graphicx}
\usepackage{float}
\usepackage[export]{adjustbox}  
\usepackage{morefloats}  
\usepackage{placeins}
\FloatBarrier
\usepackage[font=small,format=plain,labelfont=bf,up,textfont=it,up]{caption}
\usepackage{subfigure}[format=plain,labelfont=bf,up,textfont=it,up]
\usepackage{url}

\usepackage[refpage]{nomencl}
\usepackage{hyperref}

\makenomenclature
\title{Strangeness Enhancement in Proton-Proton Collisions at the RHIC Energy}
\author{Fatma. H. Sawy
\\American University in Cairo, Cairo, Egypt }
\begin{document}
\maketitle
\begin{abstract}

Strangeness enhancement is considered as a potential signature for QGP phase transition. Here we claim the observation of strangeness enhancement in proton-proton (pp) collisions at RHIC energy saying the de-confinement phase is reached. The deep study of matter phase transitions and dynamics of strangeness creation has been preformed for $K_{s}^{+}(u\bar{s})$ and $K_{s}^{-}(s\bar{u})$ mesons, $\Lambda^{0}(uds)$ baryon and for multi-strange baryons $\Xi^{0}(uss)$, $\Xi^{-}(dss)$ and $\Omega^{-}(sss)$ production. The analysis of datasets for the strange and multi-strange hadrons production is presented at $\sqrt{s} =$ 200 GeV as reported by the PYTHIA 8 event generator simulations. We preform the transverse momentum distributions of the strange and multi-strange hadrons for high and low multiplicity events. At the RHIC energy and for pp collisions, the strangeness content yield increases with the event multiplicity.
\end{abstract}


\section{Introduction}
   The strangeness enhancement as a signature of Quark Gluon Plasma (QGP) has been discussed for nearly 40 years ago\cite{alice2017enhanced} \cite{Sahoo:2021jzi}. In 1982, Rafelski and Müller proposed that QGP formation should lead to a higher abundance of strangeness per participating nucleon than what is seen in pp interactions \cite{bianchi2016strangeness} \cite{rafelski1982strangeness_article}. In 1995, the series of International Conference on Strangeness in Quark Matter first organized in Tucson, Arizona \cite{rafelski1995strangeness}. Strangeness enhancement was indeed observed for the first time by comparing central nucleus-nucleus collisions with proton-ion and pp reactions at the SPS \cite{bianchi2016strangeness}. In 2005, the Brookhaven National Laboratory (BNL) announced that forming the perfect liquid phase in nucleus-nucleus collisions \cite{jacak2010creating}. The researchers discovered spatial correlations reminiscent of the high multiplicity events attributed to QGP formation in nucleus-nucleus collisions  \cite{jacak2010creating} \cite{chang2017proton}. In 2010, the CMS collaboration spotted something unexpected in pp collisions, the experiment reported flow-like features of particle production in high multiplicity events in pp collisions \cite{sadhu2019anomalous}. In 2017,the ALICE collaboration reports the first observation of strangeness enhancement in pp collisions, a signature of QGP \cite{alice2017enhanced}. At the LHC energies, the high multiplicity events in pp collisions show features which were considered as signatures of QGP formation \cite{bianchi2016strangeness}. Since the QGP is formed in pp collisions, physicists could probe the properties of QGP in a far simpler system than the one produced in nucleus-nucleus collisions and that what we do in this paper.  \\

  Here we present the transverse momentum distributions for strange hadrons, $K_{s}^{+}(u\bar{s})$, $K_{s}^{-}(s\bar{u})$, $\Lambda^{0}(uds)$, and $\Xi^{0}(uss)$, $\Xi^{-}(dss)$, $\Omega^{-}(sss)$, for PYTHIA 8 simulated pp collisions at the center of mass energy $\sqrt{s} = 200 $ GeV. The normalized transverse momentum distributions to total number of events are provided for all of the six considered particles. The simulated data are preformed for low and high multiplicity events, from event multiplicity less than 70 to higher than 200, to compare between them. The ratio between the particle production in low and high multiplicity events for $K_{s}^{+}(u\bar{s})$, $K_{s}^{-}(s\bar{u})$, $\Lambda^{0}(uds)$, $\Xi^{0}(uss)$, $\Xi^{-}(dss)$, $\Omega^{-}(sss)$ are examined in detail.  In addition, we study the relation between the strange content yield and the event multiplicity. The interest in these studies lies in deep understanding of matter phase transitions and dynamics of strangeness creation. We show the production mechanism/s of strange and multi-strange particles. Our main objectives are introducing full complete study about QGP formation signature, the strangeness enhancement. The primary physics task of STAR experiment is to study the formation and characteristics of the QGP. Details of the STAR experiment are in \cite{abelev2008enhanced} \cite{ackermann1999star}. Since PYTHIA 8 event generator simulations cannot be performed for nucleus-nucleus collisions, we consider the high multiplicity events in pp collisions as heavy ion events. The PYTHIA 8 generator is an ideal system where the finial state interactions are not included i.e the hadrons stream freely and do not talk to each other. PYTHIA 8 is Monte Carlo event generator used in high-energy pp simulated collisions, you might be interested in setting up its parameters to study strangeness enhancement (particularly in $K_{s}^{+}(u\bar{s})$, $K_{s}^{-}(s\bar{u})$, $\Lambda^{0}(uds)$, $\Xi^{0}(uss)$, $\Xi^{-}(dss)$, $\Omega^{-}(sss)$). These settings allow you to fine-tune strangeness production in your Pythia simulations and measure the enhancement in $K_{s}^{+}(u\bar{s})$, $K_{s}^{-}(s\bar{u})$, $\Lambda^{0}(uds)$, $\Xi^{0}(uss)$, $\Xi^{-}(dss)$, $\Omega^{-}(sss)$ production.The PYTHIA 8 (version 8.309) simulation measurements are performed for $10^{5}$ pp generated events collected at $\sqrt{s} = 200$ GeV. The particle decay channels in PYTHIA 8 was set to "on" and the particle of interest was set to "off". The events that have total charges equal to zero are removed from the datasets.


\section{The Plasma and Strangeness Enhancement}
\begin{figure}[H]
\centering \includegraphics[scale=0.3]{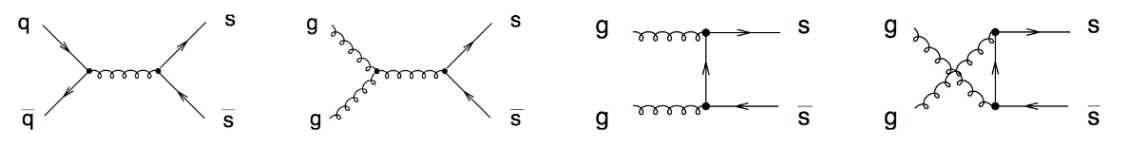}
\caption{Feynman diagrams for production of strange quarks in the QGP.}
\label{feynman_digrams_strangeness_production}
\end{figure}

  As the lifetime of QGP is ultra short, an order of $10^{-23}$ of a second, the signatures of its creation are indirect in nature. The quark flavor composition of the plasma varies during its ultra short lifetime as new flavors of quarks such as strangeness are cooked up inside. Since the initial colliding protons have no strangeness content, all new fresh strange quarks must be formed from the new plasma medium produced. The plasma evolves in time to form a liquid-like medium of quarks and gluons. The new strange quarks are created from the medium provided that there is enough time and that the temperature is high enough. T. Biro and J. Zimanyi  demonstrated that strange quarks could not be produced fast enough alone by quark anti-quark pair production. A new mechanism get alone in QGP was proposed.  Rafelski and Müller showed that the equilibrium of the strangeness yield in QGP is only possible due to gluon fusion. The gluon fusion process occurs almost ten times faster than quark - antiquark pair production. The strangeness enhancement is combined with gluon existence in QGP, in which the gluon dissolves to a pair of strange quarks rapidly as shown in \autoref{feynman_digrams_strangeness_production}. 
  
  Since $s\bar{s}$ pairs are generated together from the same medium by the same way, the both strange matter and antimatter must be originated from a common source by the same creation mechanism \cite{rafelski2006strangeness}. Since the production of strangeness occurs predominantly in thermal gluon fusion, as discussed before, the overabundant presence of strangeness is theoretical linked to presence of thermal gluons, and thus to QGP \cite{rafelski2003strangeness}. The initial hot and dense partonic matter can be described in terms of hydrodynamic and statistical models \cite{abelev2014multiplicity}. From thermodynamics point of view, the ordinary nuclear phase goes to plasma phase \cite{alice2017enhanced} at sufficient high temperature and energy density. After an initial pre-equilibrium stage, the colored strongly interacting medium produced in the collisions of the two interacting protons quickly achieves local thermodynamic equilibrium and, hence, reaches a plasma state \cite{becattini2020polarization}. The de-confined interacting QGP phase is the equilibrium state of matter at high temperature ($T$) and/or particle density ($\epsilon$) \cite{rafelski2006strangeness}. An early study explored the phase transition from hadronic matter to QGP and postulated the idea of chemical and thermal equilibrium \cite{hanafy2021strangeness}. The gluons and light quarks reach chemical equilibrium during the beginning stage of the plasma state. The heavier quark flavors such as strange typically approach chemical equilibrium in a thermodynamic evolution process. To obtain the measurable abundance of strange quarks, the lifetime of the plasma should reach after  $2 \times 10^{-23}$s. At $T = T_{ch} = 160-170$ MeV, the chemical freeze-out (chemical equilibrium) is assumed to take place immediately after hadronization of the QGP, the strangeness enhancement chemically equilibrated in plasma and the strange abundance saturation is clearly visible as well \cite{rafelski1982strangeness} \cite{huovinen2008chemical}. The chemical freeze-out temperature $T_{ch}$ and the baryon chemical potential $\mu_{B}$, describes the net baryon content of the system, mainly govern the statistical models. As the initial quark chemical potential increases, the production rate of gluons increases and the gluon equilibration rate necessarily decreases, thus leading to the little energy consumption of the system, i.e., slow cooling of the medium \cite{he2004strangeness}. The increase of the quark phase lifetime with increasing initial quark chemical potential obviously increases strangeness production \footnote{From RHIC to LHC energies, the abundances of strange particles relative to pions in nucleus-nucleus collisions do not show a significant dependence on either the initial volume (collision centrality) or the initial energy density (collision energy)\cite{alice2017enhanced}.} \cite{he2004strangeness}. The relative chemical equilibrium happens when the strangeness is originated from the quarks and gluons exchanging where the dominant mechanism of strangeness production involves gluons only present at pure QGP phase. The absolute chemical equilibrium happens when the strangeness is originated from pair production between constituents of the plasma. At $T \approx 120-160$ MeV, the strangeness abundance shows a pronounced threshold behavior.  With decreasing the temperature of the medium, the phase transition between plasma phase to hadronic phase occurs at a certain critical temperature($T_{c}$). As time evolves and the plasma medium rapidly expands and cools, remaining close to local equilibrium, the hadron gas phase is produced where the degrees of freedom are the hadronic ones, as quarks and gluons are confined \cite{abelev2014multiplicity} \cite{becattini2020polarization}. During the expansion phase of the hadron gas, collective hydrodynamic flow develops from the initially generated pressure gradients in the strongly interacting system. This results in a characteristic dependence of the shape of the transverse momentum ($p_{_T}$) distribution on the particle mass, which can be described with a common kinetic freeze out temperature parameter $T_{kin}$ and a collective average expansion velocity \cite{abelev2014multiplicity}. After a short kinetic stage of interacting hadron gas, the strongly stable particles stop colliding (kinetic freeze-out) and freely travel toward the detectors. The hadron gas is born into equilibrium and the hadrons escape from the reaction medium immediately after kinetic freeze-out \cite{becattini2020polarization}. Finally, an increase of the strange quark density demonstrate that formation of QGP in the medium and this increase results in an increase of the strange hadron production.

   From hydrodynamics point of view, the QGP is formed where the degrees of freedom are at partonic level (free quarks and gluons). The signature of the formation of QGP lies exactly on the overabundance of strangeness production in the medium. Shortly, the strange (anti)baryon and (anti)meson production processes in the two successive steps. The strange quarks are produced inside the fireball and later on, in an independent process, they hadronized from several (anti)strange quarks to (anti)baryons and (anti)meson. The energy needed to produce strange hadrons in a thermally equilibrated hadron gas is significantly higher than in the case of a QGP therefore the production of strange hadrons in hadron gas would be much more difficult \cite{noferinistrangeness}. The strangeness in the plasma phase is more abundant than in the hadronic gas phase, even if the gas is saturated. Originally, this enhancement is due to the high production rate of $gg\rightarrow s\bar{s}$ in a QGP, a process absent in the hadronic state as \autoref{feynman_digrams_strangeness_production} explained. The subsequent hadronization of these (anti)strange quarks results in a significant increase in strange hadron production, thus signaling a QGP was formed \cite{abelev2008enhanced}. At certain lower temperature, the fast expansion of the fireball will lead to super cooling which may help to understand the sudden nature of the fireball breakup that effectively described by relativistic hydrodynamic equations \cite{becattini2020polarization}. Features of strange baryon and (anti)baryon spectra demonstrate sudden breakup of the fireball into hadrons, absence of re-equilibration. The gluons attract each other which gives rise to the confinement of quarks inside hadrons. When QGP hadronized, the high availability of strange quarks and anti-quarks helps to produce matter and antimatter containing multiple strange quarks, which is otherwise rarely made. The hadronization of the QGP phase is direct, fast and sudden \cite{rafelski2006strangeness}. Many strange hadrons decay within several centimeters from the point of production \cite{rafelski2003strangeness}. Strange hadrons are naturally radioactive and decay by weak interactions into lighter hadrons on a time scale that is extremely long compared with the nuclear collision times. This makes it relatively easy to detect strange particles through the tracks left by their decay products \footnote{Strangeness enhancement has been experimentally observed for CERN SPS, RHIC and LHC energies in nucleus-nucleus collisions \cite{sahoo2019possible}}. In order to be more specific about the nature of mechanism/s of strangeness creation,
 we study the transverse momentum spectra at low multiplicity and high multiplicity events for strange mesons $K_{s}^{+}(u\bar{s})$ and $K_{s}^{-}(s\bar{u})$ and for strange baryons $\Lambda^{0}(uds)$, $\Xi^{0}(uss)$, $\Xi^{-}(dss)$and $\Omega^{-}(sss)$ in the next section. 

\section{The Transverse Momentum and Strangeness Enhancement}
\begin{figure}[H]
\centering 
\subfigure{\includegraphics[scale = 0.25]{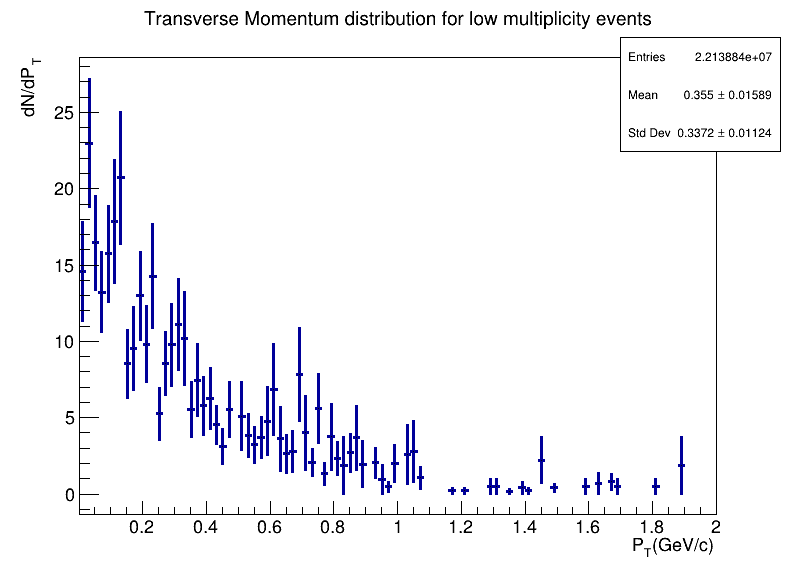}}
\subfigure{\includegraphics[scale = 0.25]{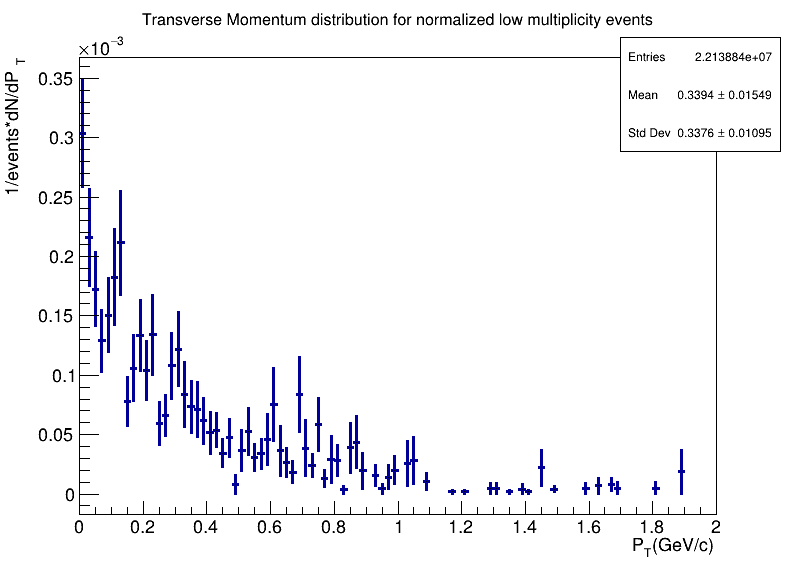}}
\caption{The transverse momentum distribution of PYTHIA 8 simulations for $k^{+}(u\bar{s})$ particles at $\sqrt{s}$ = 200 GeV for low multiplicity events, less than 70, in the left pannal and for low multiplicity events normalized to total number of simulated events in the right pannal.}
\label{k+_low_multi}
\end{figure}

\begin{figure}[H]
\centering 
\subfigure{\includegraphics[scale = 0.25]{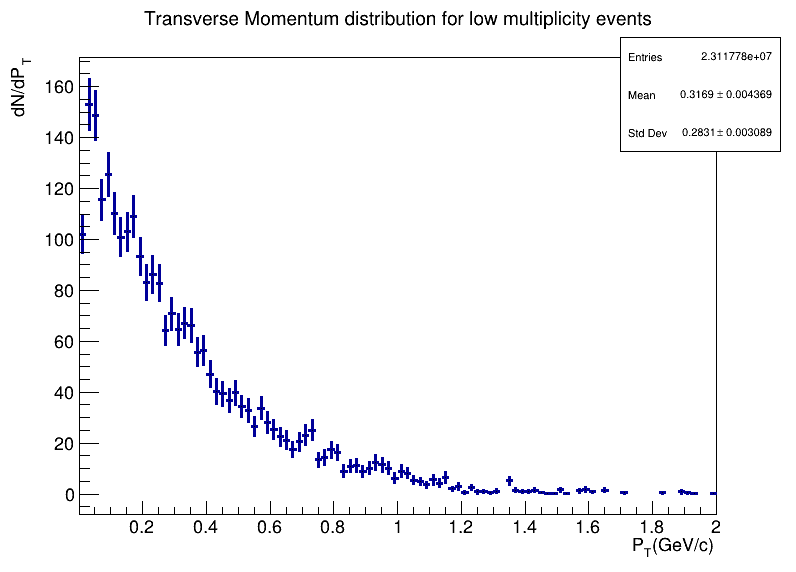}}
\subfigure{\includegraphics[scale = 0.25]{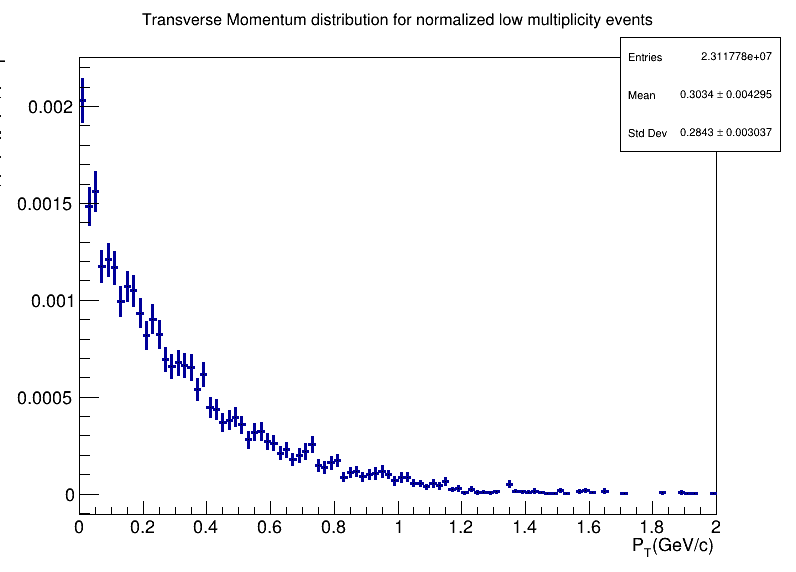}}
\caption{The transverse momentum distribution of PYTHIA8 simulations for $k^{-}(s\bar{u})$ particles at $\sqrt{s}$ = 200 GeV for low multiplicity events, less than 70, in the left pannal and for low multiplicity events normalized to total number of simulated events in the right pannal.}
\label{k-_low_multi}
\end{figure}

\begin{figure}[H]
\centering
\subfigure{\includegraphics[scale = 0.25]{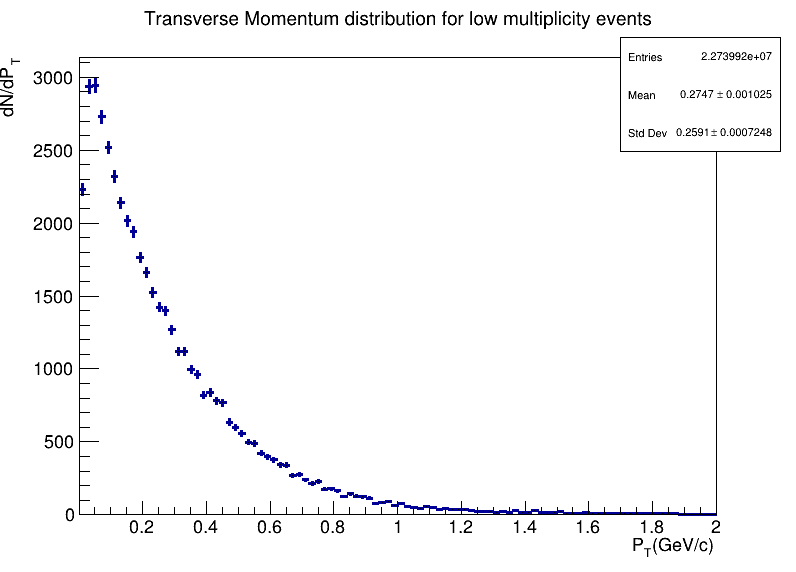}}
\subfigure{\includegraphics[scale = 0.25]{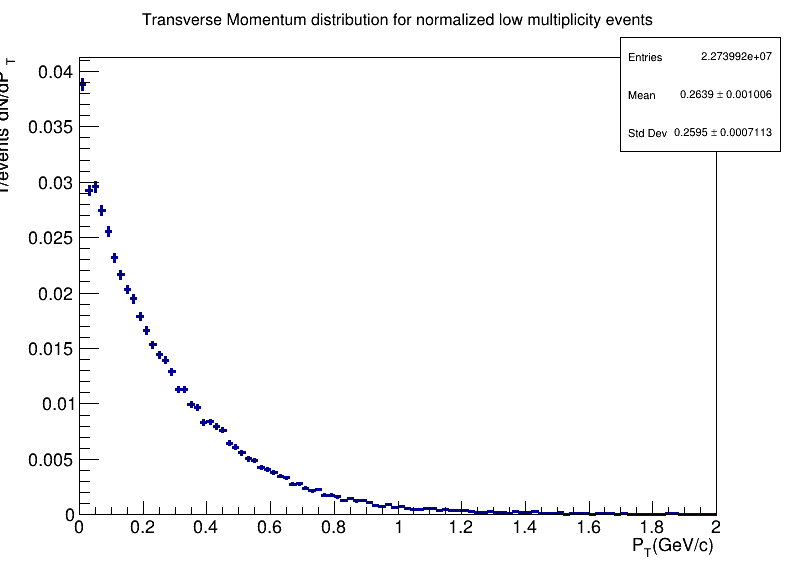}}
\caption{The transverse momentum distribution of PYTHIA8 simulations for $\Lambda^{0}(uds)$ particles at $\sqrt{s}$ = 200 GeV for low multiplicity events, less than , in left pannal and or low multiplicity events normalized to total number of simulated events in the right pannal.}
\label{lambda0_low_multi}
\end{figure}

\begin{figure}[H]
\centering 
\subfigure{\includegraphics[scale = 0.25]{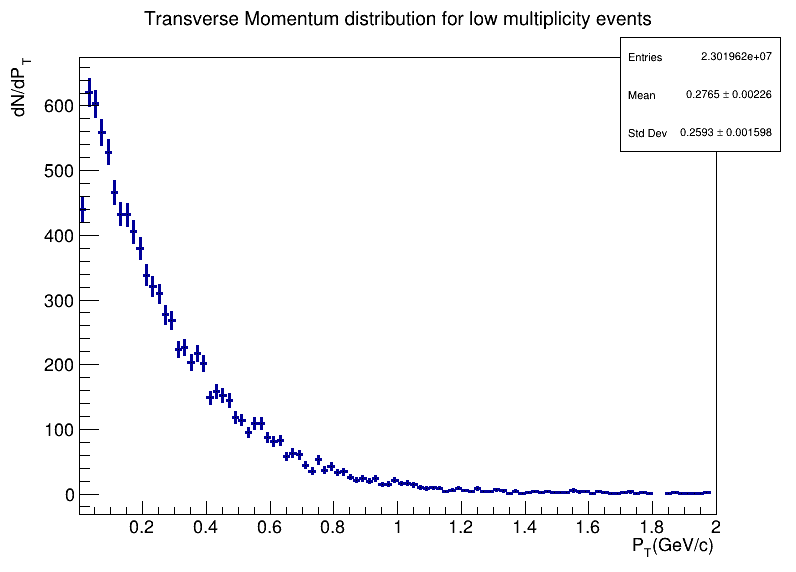}}
\subfigure{\includegraphics[scale = 0.25]{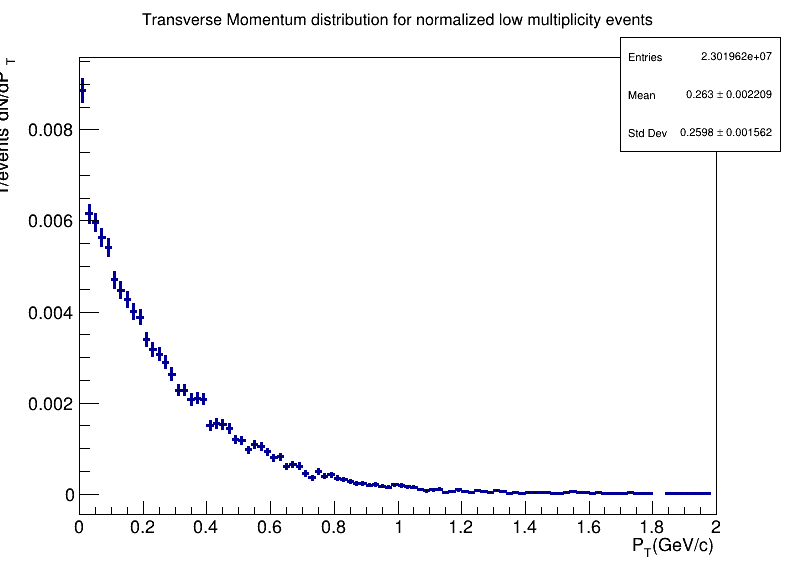}}
\caption{The transverse momentum distribution of PYTHIA8 simulations for $\Xi^{-}(dss)$ particles at $\sqrt{s}$ = 200 GeV for low multiplicity events, less than 150, in the left pannal and for low multiplicity events normalized to total number of simulated events in the right pannal.}
\label{xi-_low_multi}
\end{figure}

\begin{figure}[H]
\centering 
\subfigure{\includegraphics[scale = 0.25]{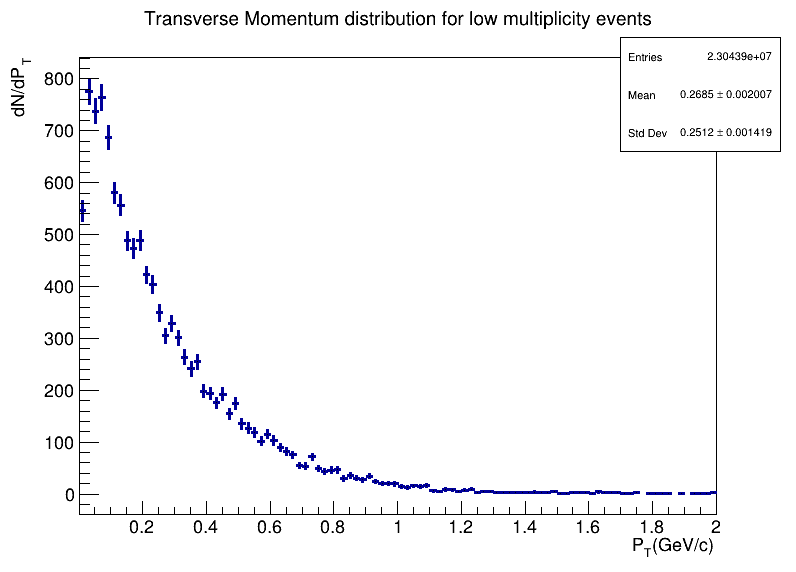}}
\subfigure{\includegraphics[scale = 0.25]{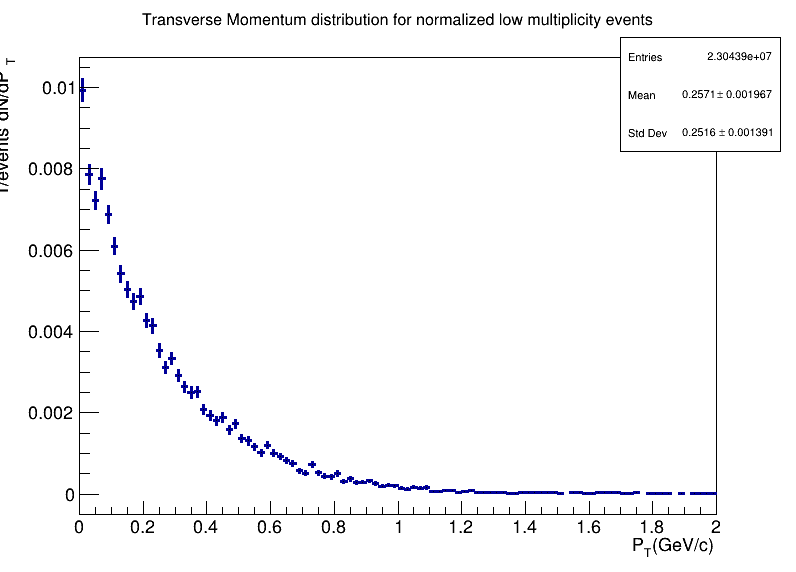}}
\caption{The transverse momentum distribution of PYTHIA8 simulations for $\Xi^{0}(uss)$ particles at $\sqrt{s}$ = 200 GeV for low multiplicity events, less than 150, in the left pannal and for low multiplicity events normalized to total number of simulated events in the right pannal.}
\label{xi0_low_multi}
\end{figure}

\begin{figure}[H]
\centering 
\subfigure{\includegraphics[scale = 0.25]{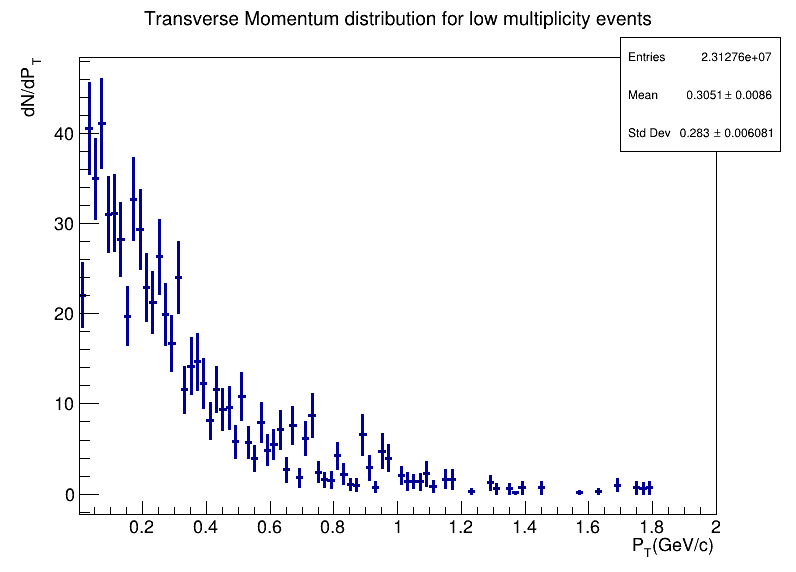}}
\subfigure{\includegraphics[scale = 0.25]{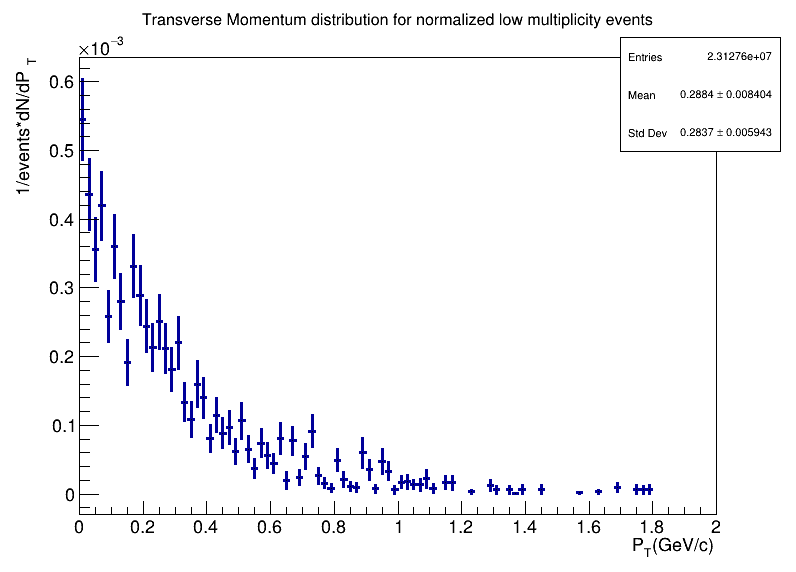}}
\caption{The transverse momentum distribution of PYTHIA8 simulations for $\Omega^{-}(sss)$ particles at $\sqrt{s}$ = 200 GeV for low multiplicity events, less than 200, in the left pannal and for low multiplicity events normalized to total number of simulated events in the right pannal.}
\label{omega-_low_multi}
\end{figure}

The strangeness enhancement in high energy nuclear interactions provides one of the tools to investigate properties of QGP \cite{Sahoo:2021jzi}. We work to study strangeness enhancement form PYTHIA 8 event generator data sets at RHIC energy for pp collisions. The measurements of transverse momentum distributions of $K_{s}^{+}(u\bar{s})$, $K_{s}^{-}(s\bar{u})$, $\Lambda^{0}(uds)$, $\Xi^{0}(uss)$, $\Xi^{-}(dss)$ and $\Omega^{-}(sss)$ are an important corner-stones of the confirm that the de-confined phase may be formed in pp collisions at $\sqrt{s} = 200$ GeV. The evidence that strangeness enhancement is obtained is coming from the study of the ratio between the strangeness production at high multiplicity events to its production at low multiplicity events for the six considered strange hadrons. The \autoref{k+_low_multi} and \autoref{k-_low_multi} prove that $K_{s}^{+}(u\bar{s})$ and $K_{s}^{-}(s\bar{u})$ can be produced in events have multiplicity less than 70 pronouncing that the strangeness is originated. 
The strangeness enhancement is observed in events having multiplicity less than  emphasized through the $\Lambda^{0}(uds)$ transverse momentum distribution, \autoref{lambda0_low_multi}. This observation is supported when the distributions \autoref{xi-_low_multi} and \autoref{xi0_low_multi} are originated proving the enhancement in $\Xi^{0}(uss)$ and $\Xi^{-}(dss)$ production for events having multiplicity less than 150. The \autoref{omega-_low_multi} establishes the overabundant of strangeness through the higher production of $\Omega^{-}(sss)$ for events having multiplicity less than 200. Since an observation of enhanced strange and multi-strange particles is called strangeness enhancement, we confirm that for events, have multiplicity less than 200,  the strangeness is produced as it is proved in \autoref{k+_low_multi}, \autoref{k-_low_multi}, \autoref{lambda0_low_multi}, \autoref{xi-_low_multi}, \autoref{xi0_low_multi} and \autoref{omega-_low_multi}. As the event multiplicity increases, the production of strange and multi-strange particle increases for these events. Additionally, we find out that for these events, as the strangeness content increases, the $dn/dp_{_{T}}$ is suppressed at hight $p_{_{T}}$. As already known, at high $p_{_{T}}$, the strangeness is produced in perturbative processes by flavor creation ($q\bar{q}\rightarrow s\bar{s}$) or ($gg\rightarrow s\bar{s}$), flavor excitation\footnote{In the flavor excitation process, a heavy sea quark belonging to an incoming hadron scatters off a quark or a gluon from the other hadron\cite{johansen2014heavy}.} ($qs\rightarrow qs$) and gluon splittings($g\rightarrow s\bar{s}$) as shown in \autoref{feynman_digrams_strangeness_production}. From intermediate to high $p_{T}$, the produced hadrons are suppressed and this effect is attributed to the energy loss of partons as they move through the hot and dense medium produced \cite{abelev2008enhanced}. The details of the energy loss mechanism guide us to understand thermodynamics of strangeness creation. At critical sharp temperature, the de-confined phase of the interacting protons is reached producing medium of interacting partons. The fireball formed from the colliding constituents originates the new quarks heavier than interacting partons through soft and hard collisions. Since $s$ quark is heavier than the lightest quarks, $u$ and $d$, the fireball starts to produce it.The up and down quarks are easily produced as quark - antiquark pairs in the hot fireball because they have small masses. The mass of strange quark, $95$ $MeV/c^{2}$, is equivalent to the temperature or energy at which hadrons dissolve into quarks. This means that the abundance of strange quarks is sensitive to the conditions of the deconfined matter phase. If the strange quark production is large, it can be assumed that deconfinement conditions were reached. As the temperature of the system is higher than the mass of a strange quark, the strange quarks and anti-quarks can be abundantly produced through several processes leading to strangeness enhancement manifesting equilibrated QGP formation. So that we had to study the low multiplicity events for strange $K_{s}^{+}(u\bar{s})$, $K_{s}^{-}(s\bar{u})$, $\Lambda^{0}(uds)$ and multi-strange $\Xi^{0}(uss)$, $\Xi^{-}(dss)$, $\Omega^{-}(sss)$ and they are reported in \autoref{k+_low_multi}, \autoref{k-_low_multi}, \autoref{lambda0_low_multi}, \autoref{xi-_low_multi}, \autoref{xi0_low_multi} and \autoref{omega-_low_multi}. From this deep study, equilibrated QGP formation could be occurred in low multiplicity events, less than 200, in pp collisions. At low $p_{T}$, non-perturbative processes dominate the production of strange hadrons through soft creation mechanism/s that is explained only by phenomenology models \cite{sawy2014}. As it notices from the $K_{s}^{+}(u\bar{s})$, $K_{s}^{-}(s\bar{u})$, $\Lambda^{0}(uds)$, $\Xi^{0}(uss)$, $\Xi^{-}(dss)$ and $\Omega^{-}(sss)$ distributions, \autoref{k-_low_multi}, \autoref{k+_low_multi} \autoref{lambda0_low_multi}, \autoref{xi-_low_multi}, \autoref{xi0_low_multi} and \autoref{omega-_low_multi}, that a high strange particle production is presented at low momentum. This observation is consist with the PYTHIA 8 event generator nature i.e the generator produces non interacting final state particles by soft creation mechanism \cite{sawy2014}.

\begin{figure}[H]
\centering 
\subfigure{\includegraphics[scale = 0.25]{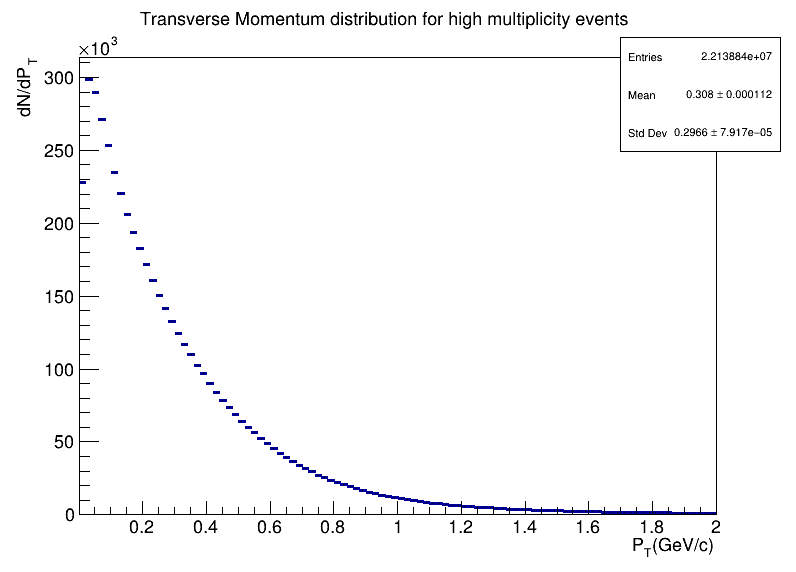}}
\subfigure{\includegraphics[scale = 0.25]{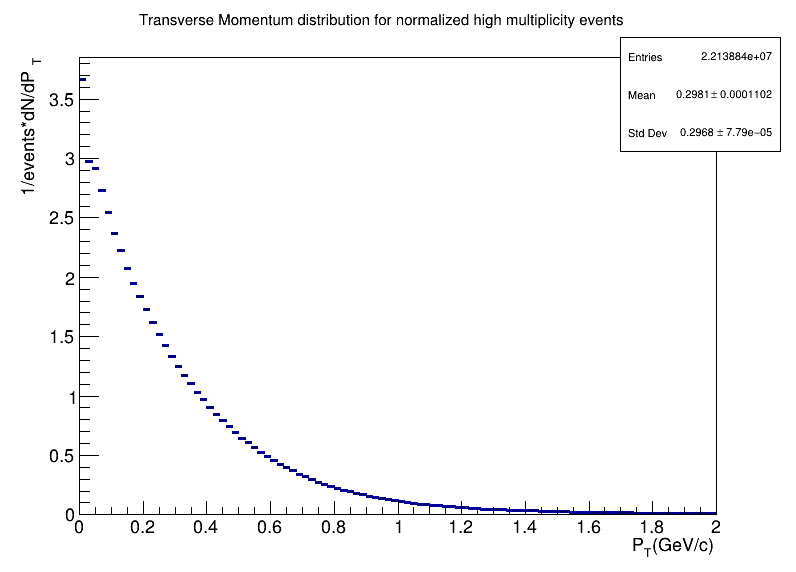}}
\caption{The transverse momentum distribution of PYTHIA8 simulations for $k^{+}(u\bar{s})$ particles at $\sqrt{s}$ = 200 GeV for high multiplicity events, higher than 50, in the left pannal and for high multiplicity events normalized to total number of simulated events in the right pannal.}
\label{k+_high_multi}
\end{figure}

\begin{figure}[H]
\centering 
\subfigure{\includegraphics[scale = 0.25]{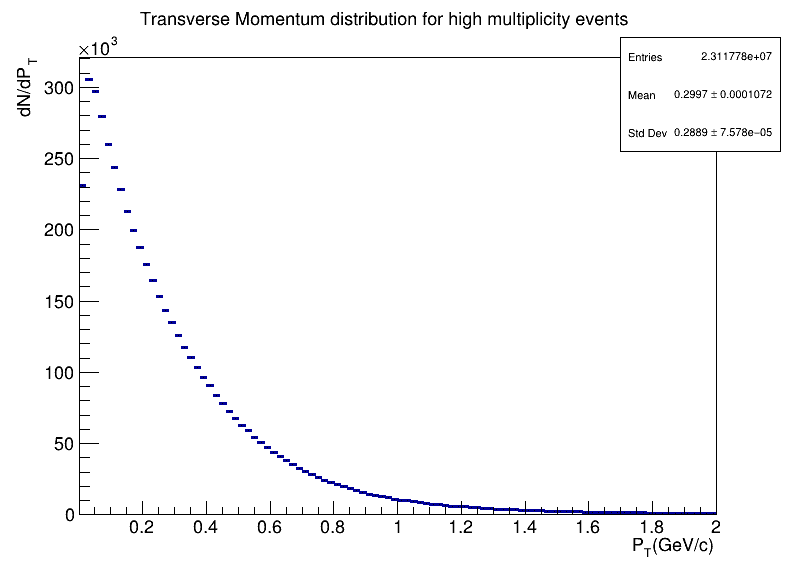}}
\subfigure{\includegraphics[scale = 0.25]{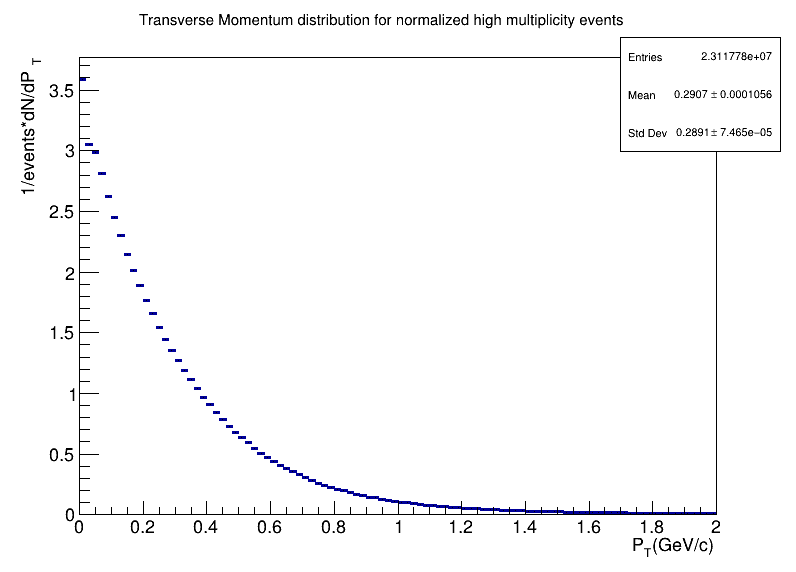}}
\caption{The transverse momentum distribution of PYTHIA8 simulations for $k^{-}(s\bar{u})$ particles at $\sqrt{s}$ = 200 GeV for high multiplicity events, higher than 50, in the left pannal and for high multiplicity events normalized to total number of simulated events in the right pannal.}
\label{k-_high_multi}
\end{figure}

\begin{figure}[H]
\centering 
\subfigure{\includegraphics[scale = 0.25]{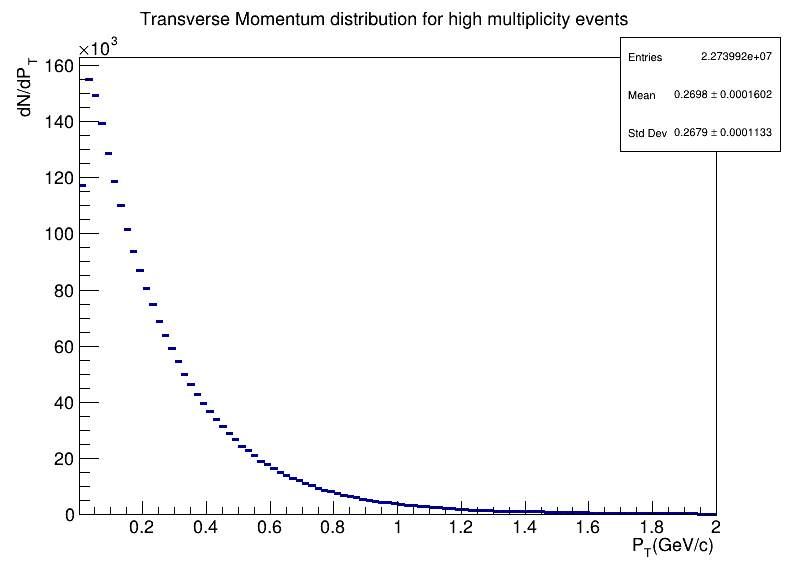}}
\subfigure{\includegraphics[scale = 0.25]{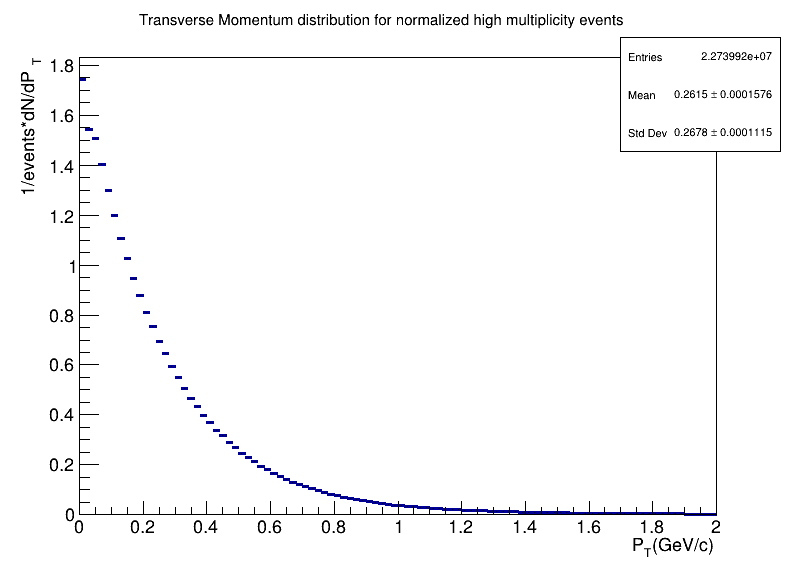}}
\caption{The transverse transverse momentum distribution of PYTHIA8 simulations for $\Lambda^{0}(uds)$ particles at $\sqrt{s}$ = 200 GeV for high multiplicity events,higher than , in left pannal and or high multiplicity events normalized to total number of simulated events in right pannal.}
\label{lambda0_high_multi}
\end{figure}

\begin{figure}[H]
\centering 
\subfigure{\includegraphics[scale = 0.25]{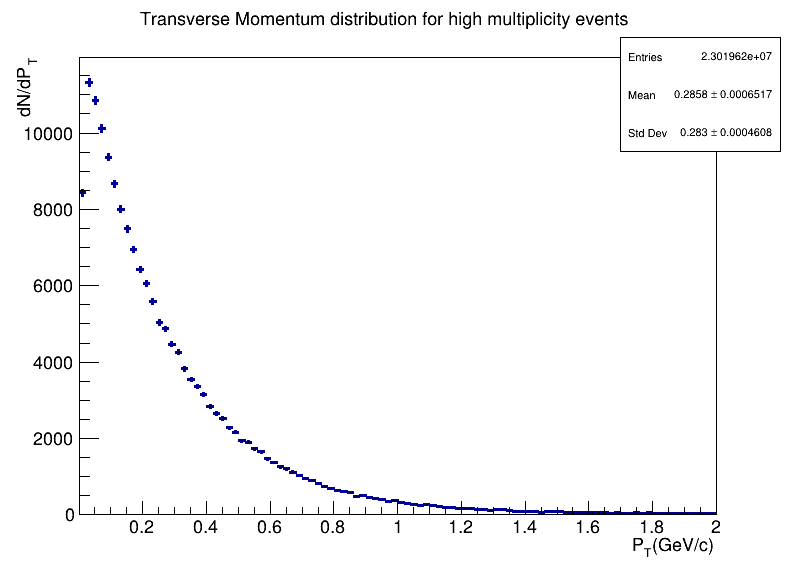}}
\subfigure{\includegraphics[scale = 0.25]{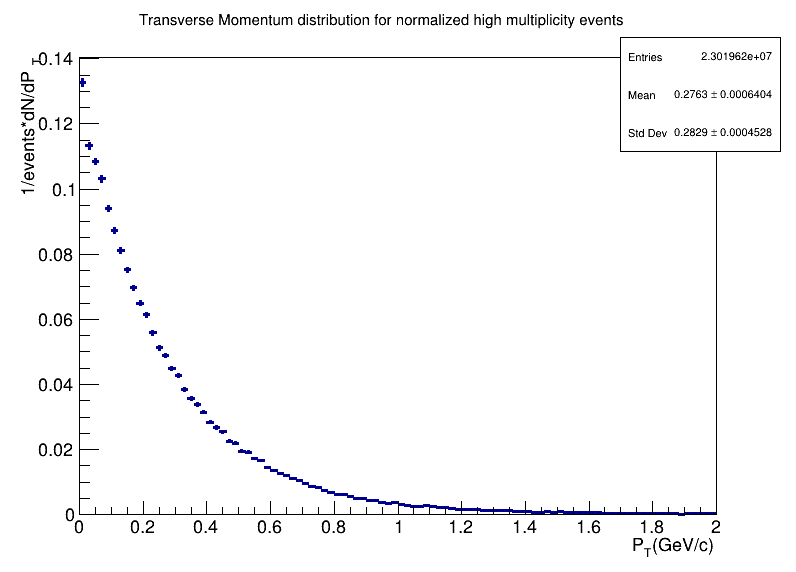}}
\caption{The transverse momentum distribution of PYTHIA8 simulations for $\Xi^{-}(dss)$ particles at $\sqrt{s}$ = 200 GeV for high multiplicity events,higher than 150, in the left pannal and for high multiplicity events normalized to total number of simulated events in the right pannal.}
\label{xi-_high_multi}
\end{figure}

\begin{figure}[H]
\centering 
\subfigure{\includegraphics[scale = 0.25]{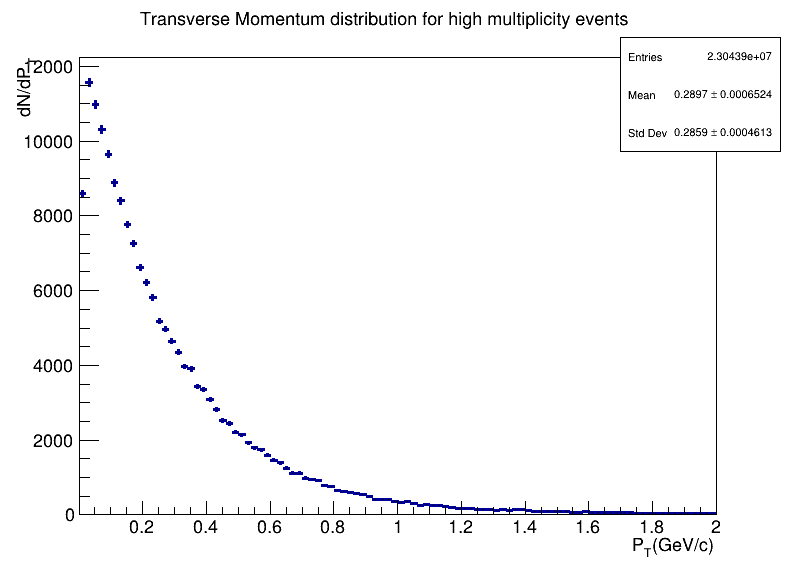}}
\subfigure{\includegraphics[scale = 0.25]{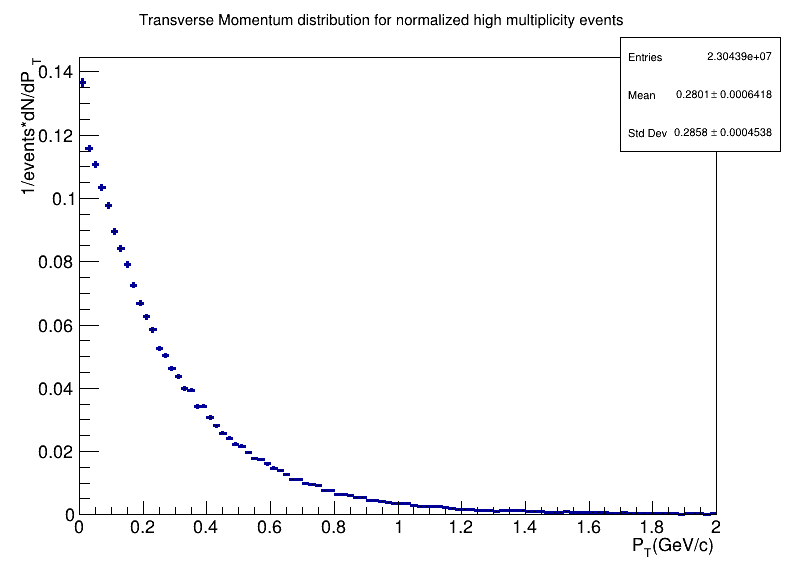}}
\caption{The transverse momentum distribution of PYTHIA8 simulations for $\Xi^{0}(uss)$ particles at $\sqrt{s}$ = 200 GeV for high multiplicity events, higher than 150, in the left pannal and for high multiplicity events normalized to total number of simulated events in the right pannal.}
\label{xi0_high_multi}
\end{figure}

\begin{figure}[H]
\centering 
\subfigure{\includegraphics[scale = 0.25]{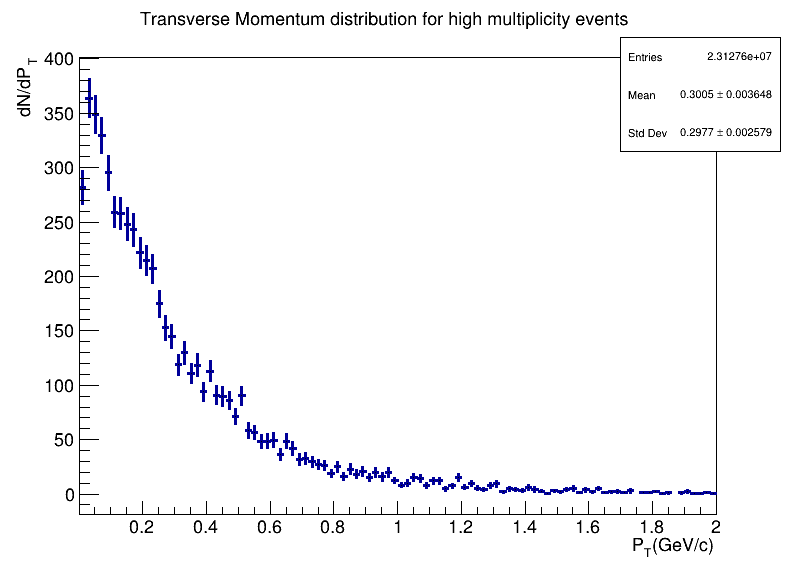}}
\subfigure{\includegraphics[scale = 0.25]{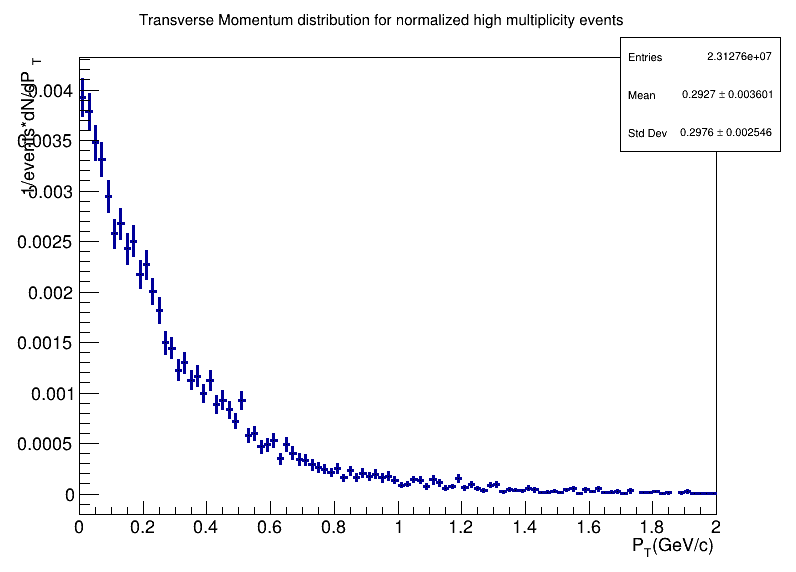}}
\caption{The transverse momentum distribution of PYTHIA8 simulations for $\Omega^{-}(sss)$ particles at $\sqrt{s}$ = 200 GeV for high multiplicity events, higher than 200, in left pannal and for high multiplicity events, higher than 200, normalized to total number of simulated events in right pannal.}
\label{omega-_high_multi}
\end{figure}

 The model of statistical and thermodynamical approximates the experimentally measured exponential momentum spectra of hadrons with a Boltzmann-like statistical distribution. With an advance of high energy collision experiments with high statistic accumulated the measured momentum spectra are found to deviate from the exponential form. Namely, at high values of particle’s transverse momentum ($p_{_{T}}$) the spectrum shows a power-law behavior. This observation is interpreted as a proof of the underlying QCD dynamics of hadronizing partonic system produced in the particle collisions.The suppression of high $p_{_{T}}$ particles identified the collective medium as a thermal medium of QGP \cite{sadhu2019anomalous}. To understand this behavior, we studied PYTHIA 8 simulations for high multiplicity events in pp collisions at STAR energy and draw the results for $k^{+}(s\bar{u})$ in \autoref{k+_high_multi}, $k^{-}(u\bar{s})$ in \autoref{k-_high_multi}, for $\Lambda^{0}(uds)$ in \autoref{lambda0_high_multi}, for $\Xi^{-}(dss)$ in \autoref{xi-_high_multi}, for $\Xi^{0}(uss)$ in \autoref{xi0_high_multi} and $\Omega^{-}(sss)$ in \autoref{omega-_high_multi}.
All these distributions confirm the suppression behavior for the $K_{s}^{+}(u\bar{s})$, $K_{s}^{-}(s\bar{u})$, $\Lambda^{0}(uds)$, $\Xi^{0}(uss)$, $\Xi^{-}(dss)$, $\Omega^{-}(sss)$ particles which manifests itself as QGP phase formation. Where the plasma present steals the high transverse momentum from the interacting fireball. The azimuthal correlations between produced particles through hard collisions are typically attributed to the formation of a strongly interacting quark gluon medium \cite{alice2017enhanced} \cite{wozniak2017study}. For mesons $K_{s}^{+}(u\bar{s})$ and $K_{s}^{-}(s\bar{u})$, \autoref{k+_high_multi} and \autoref{k-_high_multi} introduce high multiplicity events, higher than 70, normalized to total number of simulated events distributions. 

For strange $\Lambda^{0}(uds)$, we study the high multiplicity events, higher than , normalized to total number of simulated events transverse momentum distributions and draw them in \autoref{lambda0_high_multi}. 

For events having multiplicity higher than 150 normalized to total number of simulated events, we introduce transverse momentum distributions of $\Xi^{-}(dss)$ and $\Xi^{0}(uss)$ confirming overabundant of strangeness. Our results are consistent with the enhanced production of multi-strange hadrons in high-multiplicity events in pp collisions that is confirmed in \cite{alice2017enhanced}.


An even stronger signature of strangeness enhancement is the highly enhanced production of multi-strange baryons. The production of triple strange $\Omega^{-}(sss)$ is the strongest signature to pronounce QGP formation. For high multiplicity events, higher than 200, normalized to total number of simulated events distribution of $\Omega^{-}(sss)$, we draw \autoref{omega-_high_multi}.

\section{Analysis and Results}

\begin{figure}[H]
\centering
\includegraphics[scale=0.4]{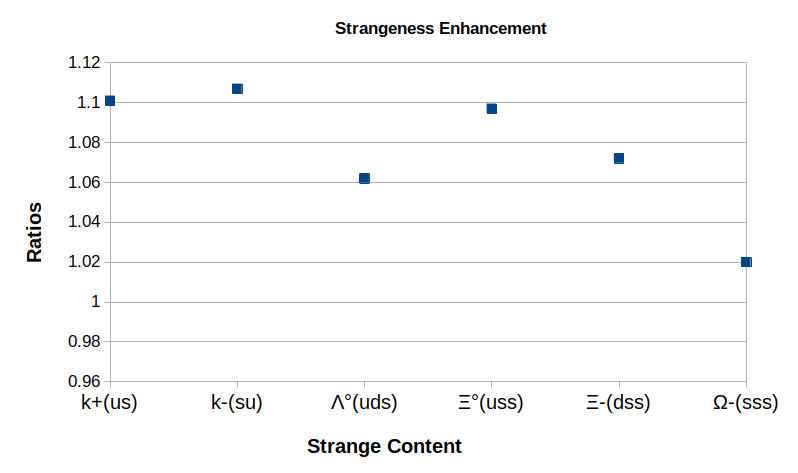}
\caption{The production ratio for $k^{-}(s\bar{u})$, $k^{+}(\bar{s}u)$, $\Lambda^{0}(uds)$, $\Xi^{-}(dss)$, $\Xi^{0}(uss)$ and $\Omega^{-}(sss)$.}
\label{strange_ratio}
\end{figure}

  The results presented here are for the considered particles at events having low multiplicity less than 200 and events having high multiplicity higher than 200. For meson $k^{+}(s\bar{u})$ and $k^{-}(s\bar{u})$, the events at low multiplicity, show lower production than the events at high multiplicity. The figure \autoref{strange_ratio} finds out the ratio between the $k^{-}(s\bar{u})$ and $k^{+}(\bar{s}u)$ production of events having multiplicity lower than 200 to their production at events having multiplicity higher than 200 is around 1.0. According to \cite{sawy2014}, the threshold energy creation of kaons is 13.5 GeV and the nature of the strangeness production comes from the color field nature which loses its energy bit by bit as excitations (kinks) of soft gluons rather than by hard single gluon radiation. This means that the soft gluons massively created hadronizing $k^{-}(s\bar{u})$ and $k^{+}(\bar{s}u)$ in low and high multiplicity events; this behavior explains the higher production of $k^{-}(s\bar{u})$ and $k^{+}(\bar{s}u)$ even the pionasatain dominates. Experimentally, the average multiplicity of keon production is 1.47 and pion production is 17.9 for full phase space at $\sqrt{s}= 200 GeV$ \cite{sawy2014}. The Wroblewski ratio between keons to pions is  about 12.17.

For $\Lambda^{0}(uds)$ particle, the production from high multiplicity events is higher than the production from low multiplicity events where the ratio around 1.062. This ratio presents that the strange production mechanism of $\Lambda^{0}(uds)$ is the same as $k^{-}(s\bar{u})$. Experimentally, the average multiplicity of lambda is 0.23 for full phase space at RHIC energy \cite{sawy2014}. The ratio between high multiplicity events to low multiplicity events is higher than 1, for $\Xi^{-}(dss)$ is 1.072 and for $\Xi^{0}(uss)$ is 1.097. The \autoref{strange_ratio} confirms that the high production in strangeness for $\Xi^{-}(dss)$ and $\Xi^{0}(uss)$ for high multiplicity events is more than low multiplicity events. The multi-strange $\Omega^{-}(sss)$ shows enhancement of strange quark that is approved in \autoref{strange_ratio}. For $\Omega^{-}(sss)$, The high multiplicity events to low multiplicity events ratio is 1.02 which strengthening the result.
\begin{figure}[H]
\centering
\includegraphics[scale=0.4]{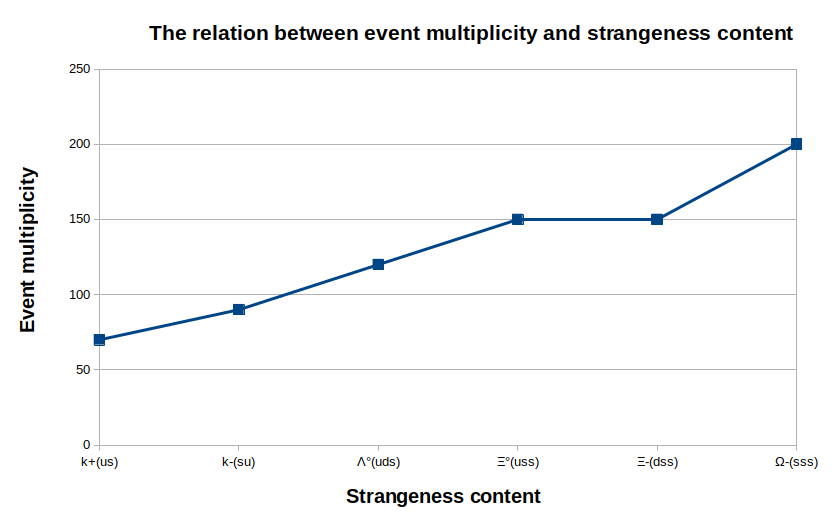}
\caption{The relation between the event multiplicity and the strangeness content.}
\label{multiplicity_limit_strangeness}
\end{figure}
We find out that in pp collisions, the strange and multi-strange particle yield increases with the event multiplicity as proved in \autoref{multiplicity_limit_strangeness}. The observations we confirmed agreed with The ALICE researchers found that in proton–lead ion collisions, the relative yield of particles that contain strange quarks increases with event multiplicity \cite{chang2017proton}. 


\section{Conclusion}
We observe the strangeness enhancement in pp collision at RHIC energy proving the claim of existence of QGP through giving a complete study about the strange and multi-strange production mechanism in $k^{-}(s\bar{u})$, $k^{+}(u\bar{s})$, $\Lambda^{0}(uds)$, $\Xi^{-}(dss)$, $\Xi^{0}(uss)$ and $\Omega^{-}(sss)$ particles. We conclude that production of the de-confined phase is obtained in high multiplicity events, higher than 200. In high multiplicity events in pp collisions, the strangeness production reaches values similar to those observed in nucleus-nucleus collisions, where a QGP is formed \cite{alice2017enhanced}.

  The suppression of the $p_{_{T}}$ spectrum at high $p_{_{T}}$ is observed for $k^{-}(s\bar{u})$, $k^{+}(u\bar{s})$, $\Lambda^{0}(uds)$, $\Xi^{-}(dss)$, $\Xi^{0}(uss)$ and $\Omega^{-}(sss)$ for high and low multiplicity events. For $k^{-}(s\bar{u})$ and $k^{+}(u\bar{s})$ meson, the low multiplicity ,less than 70, events give less strange production than the high multiplicity events, higher than 70. For $\Lambda^{0}(uds)$, the low multiplicity events, less than $120$, give less strange production than the high multiplicity events, higher than $120$. For $\Xi^{-}(dss)$ and $\Xi^{0}(uss)$ baryons, the high multiplicity events, higher then $150$, create multi-strangeness higher than low multiplicity events, less than $150$. Similarly for $\Omega^{-}(sss)$ baryons, the high multiplicity events, higher than $200$ enhance the production of multi-strange quarks than the low multiplicity events, less than $200$. There is a linear relation between the event multiplicity and the strangeness content of the produced particles as \autoref{multiplicity_limit_strangeness} demonstrated. 
\nocite{}
\printbibliography
\end{document}